\documentclass{appolb}
\usepackage{epsfig}
\usepackage{graphics}
\usepackage{graphicx}
\newcommand{\be}{\begin{equation}}
\newcommand{\ee}{\end{equation}}
\newcommand{\ba}{\begin{eqnarray}}
\newcommand{\ea}{\end{eqnarray}}
\newcommand{\bc}{\begin{center}}
\newcommand{\ec}{\end{center}}
\newcommand{\bfig}{\begin{figure}}
\newcommand{\efig}{\end{figure}}

\def\RAPGAP{{\sc Rapgap}}

\newcommand{\alphas}{\alpha_s}

\newcommand{\kt}{k_{t}}

\newcommand{\gap}{\stackrel{>}{\sim}}

\def\PO{\ensuremath{I\! \! P}}
\def\RE{\ensuremath{I\! \! R}}
\def\xpom{\ensuremath{x_{\PO}}}

\begin{document}
\title{The Diffractive Interactions Working Group Summary%
}
\author{H. Jung$^1$, R. Peschanski$^2$, C. Royon$^3$
\address{  
$^1$ Department of Elementary~Particle~Physics, Lund~University, 
S-22100~Lund,~Sweden \\
$^2$ Service de Physique Th\'eorique,
CE-Saclay, F-91191 Gif-sur-Yvette Cedex, France\\
$^3$ Service de Physique des Particules,
CE-Saclay, F-91191 Gif-sur-Yvette Cedex, France}
}

\maketitle
\begin{abstract}
Diffractive interactions represent a lively domain of investigations, as 
confirmed by the progresses reported during the conference. We summarize
the diffractive interactions session and put the new experimental data (section 
{\bf 1}), developments in modeling diffraction (section {\bf 2}) and  the 
theoretical relations with Quantum Chromodynamics (section {\bf 3}) in 
perspective.
\end{abstract}

\section{HERA and Tevatron on Diffraction}

The basis for the  investigations on diffraction is provided by a large and 
interesting set of new experimental data. Indeed, the phenomenon of ``hard'' 
diffraction, where a hard probe has been proven to be consistent with a 
diffractive process, came first as an experimental surprise both at 
hadron-hadron ($Sp\bar pS,$ 88', Tevatron 95') and  lepton-hadron (HERA, 94') 
colliders. 
It is thus to be remarked that  new interesting results have been provided this year by 
thorough  experimental studies at HERA and Tevatron, leading also to a preview 
on what can be expected from HERA, Tevatron (Run II) and LHC on diffraction in 
the future.

\subsection{Diffractive structure function measurements at HERA and QCD 
fits}

\begin{figure}
\begin{center}
\mbox{\epsfig{figure=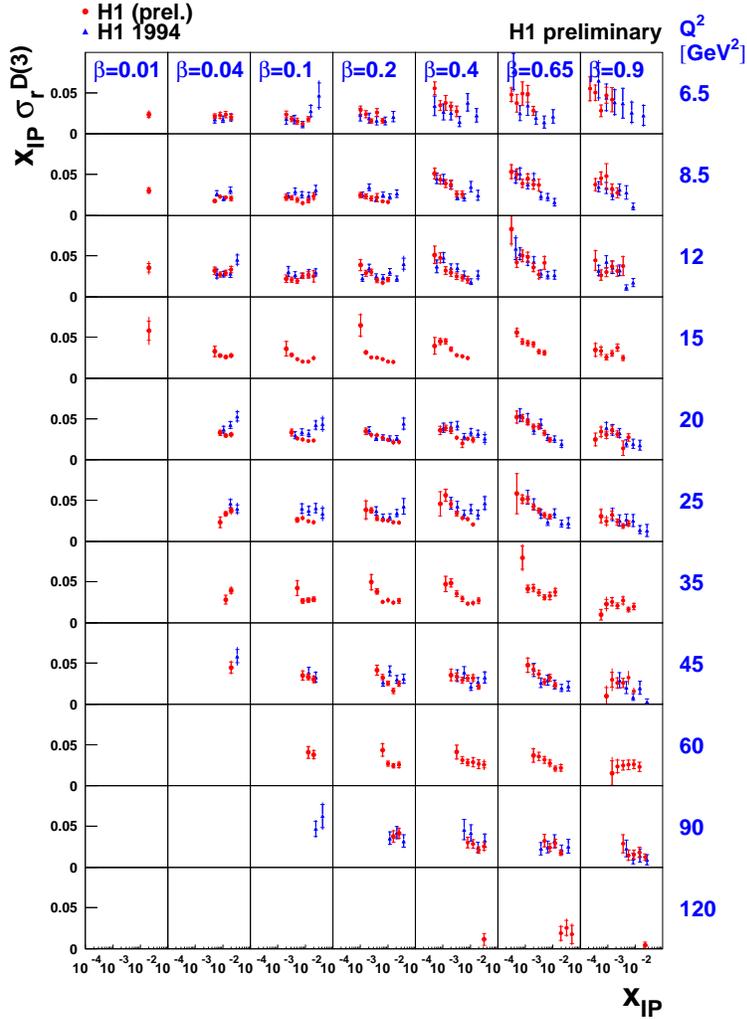,height=140mm}}
 \end{center}
 \caption{Comparison on the new (1997) and old (1994) measurements of the proton
 diffractive structure function from the H1 experiment.}
 \label{94vs97}
\end{figure}

The H1 collaboration \cite{laycock} presented at this conference a new
measurement of the diffractive structure function employing the rapidity gap method
using the 1997 data representing 10.6 pb$^{-1}$ which is 5 times more luminosity
than for the previous measurement. In Fig.~\ref{94vs97} is
presented the new $F_2^D$ measurement together with the old data of 1994.
The kinematical domain is 6.5 $<$ Q$^2 <$ 120 GeV$^2$, 0.01 $< \beta <$ 0.9
and $10^{-4} < x_{\PO} < 0.05$. Globally a good agreement between the 
old and the
new measurement is observed. A few bins show large differences which explain the differences
in the QCD NLO DGLAP fit as we will see in the following. A good agreement
between the rapidity gap measurement and the Forward Proton Spectrometer data
has also been shown \cite{laycock}.
A Regge fit of the new
data leads to a new measurement of the pomeron intercept
$\alpha_{\PO}(0) = 1.173 \pm 0.018 (stat) \pm 0.017 (syst) ^{+0.063}_{-0.035} 
(model)$
in good agreement with the previous measurement. The growth of $\alpha_{\PO}(0)$
as a function of $Q^2$ is slower for diffractive events than for inclusive ones 
\cite{laycock}. The intercept of the secondary reggeon was fixed to 0.5,
consistent with the previous value~\cite{schilling-private}. 
\begin{figure}
\begin{center}
\mbox{\epsfig{figure=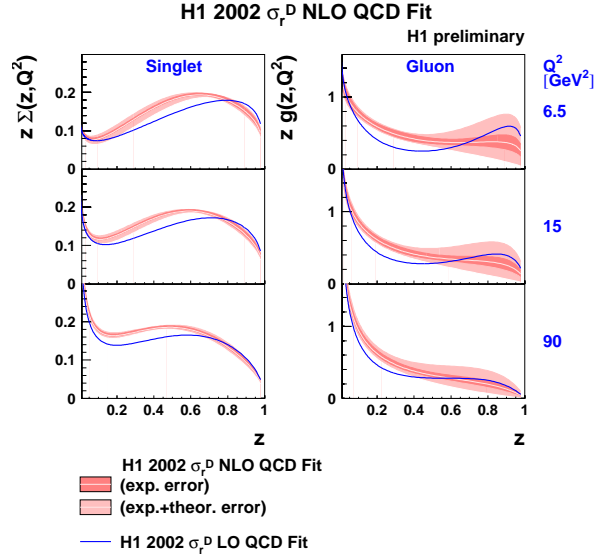,height=80mm}}
 \end{center}
 \caption{Quark and gluon density in the pomeron obtained with a NLO and a LO 
DGLAP QCD
 fit of the H1 diffractive data.
 }
 \label{fitnlo}
\end{figure}

\begin{figure}
\begin{center}
\mbox{\epsfig{figure=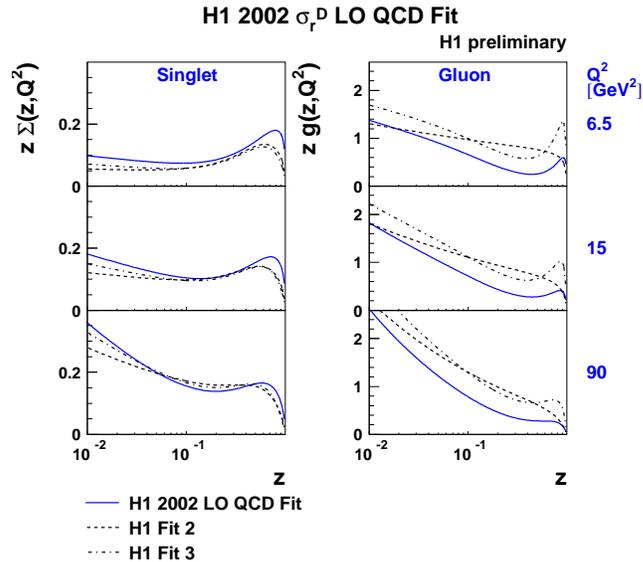,height=80mm}}
 \end{center}
 \caption{Quark and gluon density in the pomeron obtained with a LO DGLAP QCD
 fit of the H1 diffractive data, compared with the result using 1994 data.
 }
 \label{fitlo}
\end{figure}

The H1 collaboration has also presented a new NLO DGLAP QCD fit of the new 1997
data. The data used in the fit cover $x_{\PO} <0.05$, $0.01 \le \beta \le 0.9$ and
$M_X > 2$ GeV. The latter cut is applied to justify a leading twist approach.
Two set of data are used in the fit: the new H1 preliminary data presented at
this conference ($6.5 \le Q^2 \le 120$ GeV$^2$), and the higher $Q^2$ data
($200 \le Q^2 \le 800$ GeV$^2$) \cite{schilling}. The quark and gluon
distributions together with their errors are given in Fig.~\ref{fitnlo}. 
The gluon is found to be dominant. The comparison at LO with the previous data
is given in Fig.~\ref{fitlo} and shows differences for the quark density at low
values of $\beta$, and for the gluon density over the full range where the gluon
is found to be smaller than before. Taking into account the uncertainties
(especially of the old fits) the agreement is still reasonable. As we will see
in the following, this new QCD fit will have important consequences on 
diffractive final state predictions and on comparison with Tevatron data.
Integrating the parton distributions over the measured kinematic range
leads to a determination of the momentum fraction of the color singlet exchange
carried by gluons, which is found to be $75 \pm 15$\% at $Q^2=10$ GeV$^2$
\cite{schilling}.

\subsection{Diffractive and total cross sections}

The ratio of diffractive to the total cross section was known to be energy
independent, if investigated as a function of $M_X$~\cite{capua}.
In this context, progress has been made in understanding also
the leading neutron cross sections, as they can be reasonably well described by
the one-pion-exchange mechanism and in terms of the pion structure
function~\cite{borras}.
\par
In vector-meson ($V$) production the energy dependence of the ratio
$\sigma(ep \to e V p)/\sigma_{tot}(ep)$  was investigated~\cite{levy}.
The energy dependence of $\rho$ meson production is found to be the same as for 
the inclusive cross section for values of $ 0.15 < (Q^2 + M^2)/4 < 6.9$~GeV$^2$,
whereas $J/\psi$ production shows a steeper energy dependence in the same range
of $(Q^2 + M^2)/4$. In models, where diffraction is directly related to 
non-diffraction, a steep energy dependence is expected, as observed in the case
of $J/\psi$ production. This observation supports the picture that the 
pQCD cross section for vector-meson production goes like $~
\left[xG(x,Q^2)\right]^2$, with $xG(x,Q^2)$ being the standard proton gluon
density. The observation of no energy dependence in case of the $\rho$ meson and
also for inclusive diffraction implies that the contribution of hard and soft
processes is different in diffraction and non-diffraction.
\par
When comparing the ratio with theoretical expectations one should keep
in mind, that the inclusive cross section is measured at $t=0$ whereas in
diffraction normally a range in $t$ is integrated over, which due to shrinkage
in the soft diffractive part, can easily lead to a change of the energy slope of
the order of 20 \%. If, however, even with a larger lever arm in energy, this
ratio stays constant as a function of the energy, much can be learned about the
interplay of soft and hard interactions.

\subsection{Vector meson production}

At this conference,
new measurements of vector meson ($\rho$, $J/\psi$)
production both at large $|t|$ and at large
$Q^2$ have been presented~\cite{klimek,janssen}, showing a behavior typical for
hard scattering processes, if $|t|$ or $Q^2$ is large enough. In photoproduction
also $\psi(2S)$ mesons have been investigated~\cite{brown} and analyzed in terms
of the dipole picture which shows sensitivity to the different wave functions 
of
the $J/\psi$ and  $\psi(2S)$ mesons. Exclusive $\rho$ production has been
studied at HERMES~\cite{airapetian} 
in terms of the coherence length of the photon.
The vector meson studies have also been extended to cover the lightest vector
meson, the photon, via the DVCS process~\cite{stamen}. The measurements can be
reasonably well described in terms of the approach using generalized parton
distributions~\cite{freund}. 
\par
Instead of vector mesons with $C=-1$, final states with $C=+1$ can be produced by
Odderon exchange. However, no experimental signal for the $C=+1$ states could
be observed~\cite{berndt}.  In diffractive dissociation
in photoproduction the pomeron intercept
$\alpha_{\PO}=1.127\pm0.004\pm0.025\pm0.046$ has been obtained from a triple
Regge analysis~\cite{heremans}, which agrees within the errors with the value 
obtained in diffractive DIS.

\subsection{Diffraction at Tevatron}

\subsubsection{Hard diffraction with rapidity gaps}

\begin{figure}
\begin{center}
\mbox{\epsfig{figure=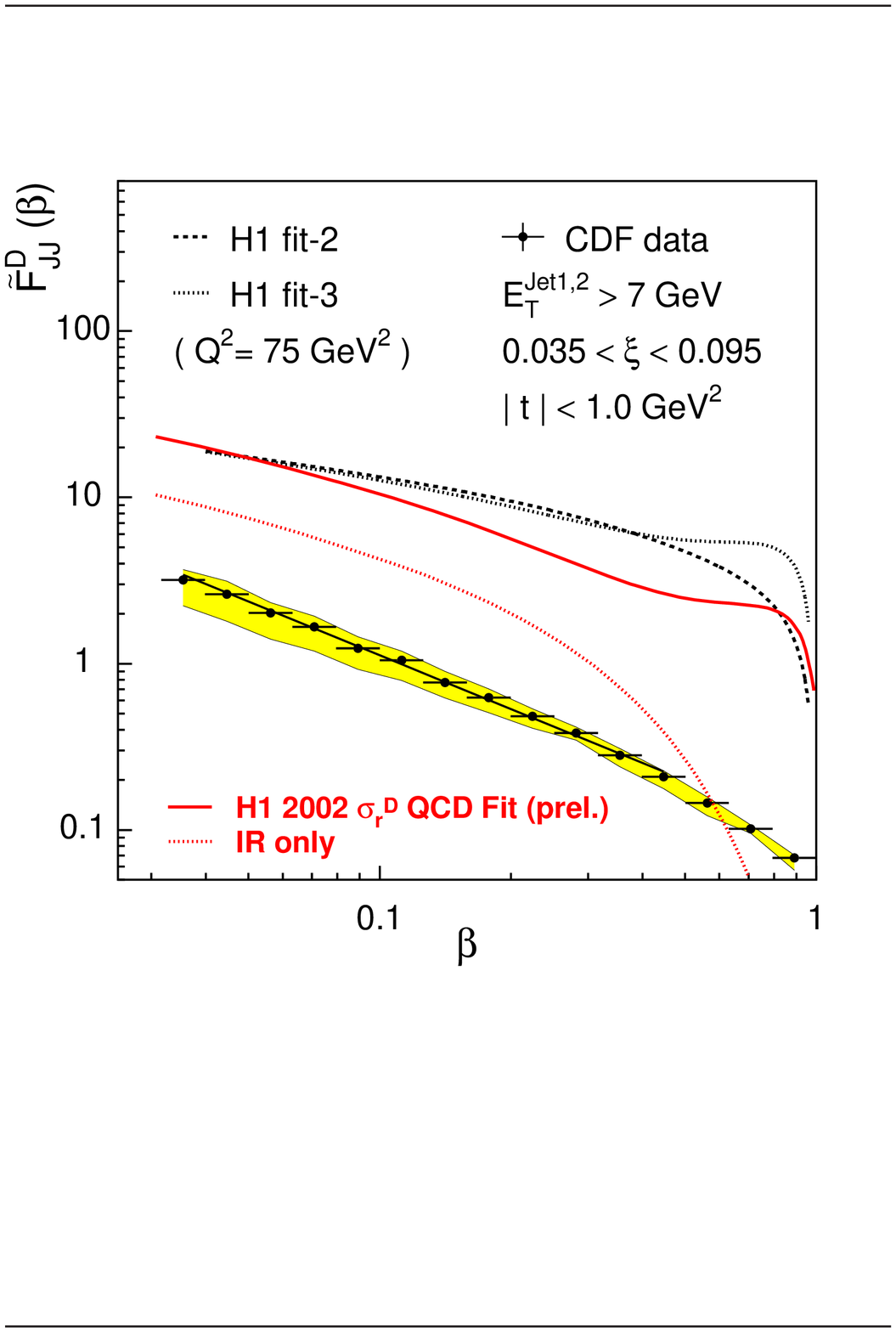,height=60mm,clip=}}
 \end{center}
 \caption{Comparison of the CDF measurement of the antiproton diffractive
 structure function with the extrapolation of the H1 measurement using a DGLAP
 NLO QCD fit. We note a discrepancy in normalization but the shapes seem to be
 similar.}
 \label{compare}
\end{figure}

Using the D\O\ and the CDF detectors, 
events produced in $p \bar{p}$ collisions with large
rapidity gaps have been investigated at the Tevatron Run I for two center-of-mass
energies of 630 and 1800 GeV \cite{simak,terashi}. The D\O\ and CDF
collaborations have studied the forward and central diffractive dijet 
productions. The D\O\ collaboration is requiring two jets with $E_T>12$ or
15 GeV in two different pseudorapidity region $| \eta | <1$ (central region)
and $ | \eta | >1.6$ (forward region). The fraction of diffractive events 
in the two kinematical regions for the two center-of-mass energies have been
compared to different theoretical predictions. The hard and flat gluon rates 
predicted by the Monte Carlo are
higher than in the data, and the fact that the rates are higher for forward jets
than for central ones. This indicates that the data are more compatible with a
combination of a soft $(\sim (1-\beta)^5)$ and hard gluon $(\sim (1-\beta))$ density 
function in the pomeron. A pomeron made of quarks is also compatible
with the data \cite{simak}. Contrary to the D\O\ collaboration who corrects
the Monte Carlo prediction and not the data, 
the CDF collaboration chose to correct the
experimental diffractive rates for ``rapidity gap acceptance", defined as the
ratio of events with $\xi$ less than 0.1. This explains most of the differences 
in
diffractive rates published by both experiments. The CDF collaboration has
measured the diffractive rates for the Dijet$+$Gap, Jet$+$Gap$+$Jet, diffractive
$b$, and $J/ \Psi$ production, and they are all in the 1\% range \cite{terashi}.
Since the measured single diffractive processes have different sensitivities
to the quark and gluon content of the pomeron, it is possible to determine the
gluon fraction of the pomeron $f_g$ which has been found to be $0.54^{+0.16}
_{-0.14}$ to be compared with the HERA result $75 \pm 15$\% at $Q^2=10$
GeV$^2.$~\cite{schilling}

The new result by the D\O\ collaboration consists in
the observation of diffractively produced $W$ and $Z$. The diffractive $W$ rate
is $0.89^{+0.20}_{-0.19}$\% (forward: 0.64\%, central: 1.08\%), and the
diffractive $Z$ rate is  $1.44^{+0.62}_{-0.54}$\%. This is the first time that an evidence
of diffractive $Z$ production is shown. The CDF collaboration also reports
evidence for diffractive $W$ production, and their rates is $1.15 \pm 0.55$\%.

\subsubsection{Hard diffraction with tagged protons or antiprotons}

The CDF collaboration installed roman pot detectors at the end of Run~I to be
able to tag the antiprotons in the final state. In leading order, the ratio
$R^{SD}_{ND}$ of single diffractive to non-diffractive dijet production rates is
equal to the ratio of diffractive to non-diffractive structure functions of the
antiproton. The measurement of this ratio and the knowledge of the antiproton
structure function leads directly to a measurement of the diffractive antiproton
structure function \cite{terashi} which can be compared directly to the HERA
measurement. Fig.~2 of Ref.\cite{terashi} shows the measured diffractive
structure function in the kinematic region $|t| <$1 GeV$^2$, 0.035 $< \xi
<$ 0.095, and $E_T^{jet1,2}>$7 GeV. The comparison between the HERA
extrapolation to the CDF kinematical domain and the CDF results is given in 
Fig.~\ref{compare} and shows a discrepancy in normalization of a factor 7 to 8
between both experiments. However, contrary to previous results based on
previous H1 measurements, the shape of both distributions seems to be quite
close \cite{terashi}. It seems that these results could be interpreted in term
of a survival gap probability which does not seem to be kinematically
dependent. The knowledge of the secondary reggeon structure function is also
not constrained from HERA (the pion structure function is assumed) and can surely influence
the comparison between the HERA and Tevatron results. The understanding of
diffractive results from HERA and Tevatron together is clearly a theoretical and
experimental challenge in the next future, and is fundamental if one wants to
understand what diffraction is. This is also important for extrapolating  Tevatron
results to LHC as we will discuss in the following.

The D\O\ collaboration has also installed roman pot detectors which will allow 
to
increase notably the possibilities for diffractive measurements at Run II.
The D\O\ Forward Proton
Detector (FPD)~\cite{FPD} consists of momentum spectrometers which
allow to measure the energies and angles of the scattered proton and
antiproton in the beam pipe. 
Tracks are measured using scintillating
fiber detectors located in vacuum chambers positioned in the
Tevatron tunnel 20--60 meters upstream and downstream of the
central D\O\ detector. Results using these new detectors are expected for the
next DIS conference.

\subsubsection{Double-gap soft diffraction}

The CDF collaboration presented a study of $p \bar{p}$ collisions with a leading
antiproton and a rapidity gap in addition to that associated with the 
antiproton \cite{goulianos}. They find that the two-gap to one-gap event ratio
is larger than the one-gap to no-gap ratio. It means phenomenologically that the
``price" to pay to get a second gap is smaller than the ``price" to get one gap.
The data are in agreement with the predictions of the gap renormalization
approach \cite{goulianos}.

\subsection{Future of diffraction}

\subsubsection{Diffractive Higgs production at LHC}

At this conference, the more recent approaches giving predictions for
diffractive Higgs cross sections were discussed \cite{albert}. We can
distinguish between proton-based \cite{martin}, pomeron-based \cite{maarten} and
soft color interaction based models \cite{nicusor}. These models are compared in
detail in Ref \cite{albert}, and we will only summarize the main results.
The proton-based models \cite{martin} do not assume the existence of the pomeron 
and 
lead to a specific signature where we have the decay products of the Higgs boson
in the main detector, the tagged protons in the roman pot detectors and nothing
else (exclusive model). This model also predicts the dijet cross sections
and thus can be tested directly already at
Tevatron. In particular, it leads to a peak in the dijet mass distribution at
high values and the Tevatron experiment will be a direct test of this model.
The pomeron-based models \cite{maarten} (inclusive models) assume a pomeron made 
of quarks and 
gluons. These models produce a Higgs in the main detector together with the
pomeron remnants, and two tagged protons in the roman pot detectors. It is
possible to reconstruct precisely the Higgs mass provided one can measure the
amount of energy lost in the pomeron remnants \cite{maarten}. The third set of
models based on soft color interaction \cite{nicusor} assumes that diffraction 
is due to color
rearrangement in the final state, and thus does not assume the existence of the
pomeron. As shown in Table \ref{higgs}, these models lead to different
predictions of the diffractive Higgs production at Tevatron and LHC.
All models give low cross sections for the Tevatron, and looking for diffractive
Higgs seems to be difficult except for the supersymmetric Higgs, which might
increase the cross section notably. At LHC, all models except the soft color
interactions give quite high cross section, but still show big differences.

It will be quite important to distinguish between the different models in the
near future at the Tevatron. All models can give predictions for diffractive
dijet or diphoton production at the Tevatron, and the forthcoming measurements in
the next years will allow to distinguish experimentally between the models and
to make better predictions for LHC. Another issue is the energy dependence of
the survival gap probability which needs to be understood theoretically if one
wants to make predictions for LHC \cite{bialas}.

\begin{table}
\begin{center}
\begin{tabular}{|c||c|c|c|c|} \hline
  & (1)     & (2)   & (3) & (4) \\ \hline \hline
 H $\sim$ 120 GeV, Tevatron & 0.3     & 3.   & 22 & 1.2 10$^{-3}$  \\
 H $\sim$ 120 GeV, LHC & 14. & 3600.   & 3219. & 0.2  \\
 H $\sim$ 160 GeV, LHC & 5.5 & 1460.   & 2100. & ? \\ \hline
\end{tabular}
\end{center}
\caption{Number of events for 10 fb$^{-1}$ for diffractive Higgs production for 
different
models (exclusive Higgs production (1), inclusive Higgs production in
the factorizable (2) and non factorizable (3) cases, soft color interaction
model (4)).}
\label{higgs}
\end{table}

\subsubsection{Prospects on experimental facilities}

After more than 10 years after the first results on hard diffraction, the
picture of high energy processes becomes clearer. In many aspects diffraction is
similar to non-diffractive scattering, and only now with precise measurements,
deviations from the simple expectations are observed. However, more
investigations are needed to substantiate such observations. With the upgraded
HERA machine a substantial increase in luminosity is still expected, and  
with the
installation of the new  very forward proton spectrometer (VFPS) exciting new
results can be expected~\cite{newman}. 
\par
The understanding of diffraction seems to act as a key to the
understanding of the structure of the proton, be it in terms of confinement or
in terms of high parton densities and self interaction leading to saturation, or
simply in terms of the small $x$ evolution of parton densities. Given the importance of
understanding QCD as a whole, several future collider options
are presently discussed
where these and other questions could be answered~\cite{caldwell} 
(see Tab.~\ref{future}).
\begin{table}
\begin{tabular}{|c | c |c |c |}
\hline 
collider & energies $[GeV]$ & $\sqrt{s}$  $[GeV]$ & lum $[/sec/cm^2]$\\ \hline 
 EIC & $E_e= 3-10$,  $E_p\sim 30-250$  &
 $\sim 20-100$ & $\sim 10^{33} - 10^{34}$ \\\hline
HERA III & $E_e \sim 30 $,  $E_p\sim 820-920$  &$300-330$& \\\hline
 HERA & $E_e= 250-500$ $E_p\sim 1000$ & $1000-1414$ & $\sim 10^{30}$ \\\hline
 \end{tabular}
 \caption{Possible future collider options at {\sc Electron Ion Collider} (EIC),
 HERA~III and THERA.
 \label{future}}
\end{table} 

\section{Modeling Diffraction}

Several theoretical approaches to calculate total cross sections in
diffraction exist. However, modeling the hadronic final state of diffractive 
processes
in detail  is much more complicated compared to non-diffractive scattering,
because the color connection of the produced partons might 
influence on the production of the rapidity gap or as in the case of charm, 
the $c \to D^*$ fraction might be different compared to the inclusive case,
since also $c \bar{c}$ bound states, like $J/\psi$ might be produced.
In addition 
multi-gluon emissions in the QCD cascade, initial and final state parton showers
or color dipole emissions, have to be modeled properly, since they can easily
affect the hadronic final state as well as the formation or
destruction of the rapidity gap. 

\subsection{Models}

The following three approaches, applicable to diffractive scattering 
in $ep$ and $p\bar{p}$ processes, are included in full hadron level 
Monte Carlo programs:
\begin{itemize}
\item  
The {\it resolved pomeron model} (Fig.~\ref{diagram}(a.)) either based on
the Ingelman-Schlein ansatz with the additional assumption 
of Regge factorization (\PO\ - flux), or in the more general formulation 
applying the factorization theorem of Collins.
\item 
The {\it perturbative 2-gluon} picture~\cite{kyrieleis} (Fig.~\ref{diagram}(b.)) either in its realization 
in the wave function approach, in the dipole model
or in the $k_t$-factorization approach.
\item  
The {\it soft color interaction model}~\cite{nicusor} (Fig.~\ref{diagram}(c.)) and its extension 
to the generalized
area law model.
\end{itemize}
Please note, that we have chosen 
for practical reasons short names to identify the different approaches.

\subsubsection{Resolved pomeron model}
\begin{figure}
{\scalebox{0.43}{\includegraphics*{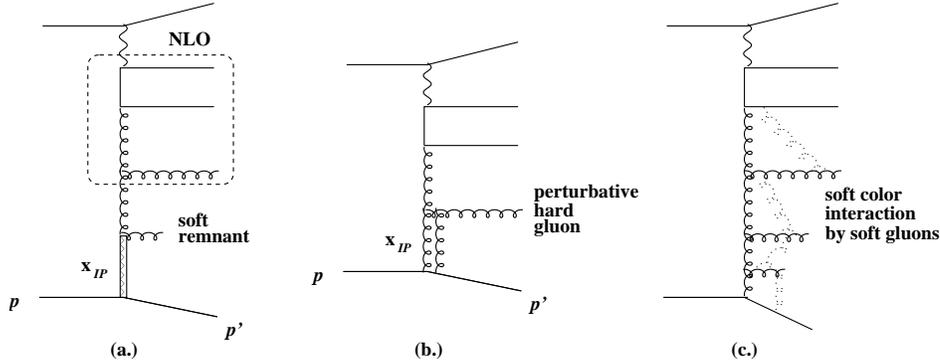}}}
\caption{Schematic diagrams for diffractive $ep$ scattering in the {\it resolved
pomeron model} $(a.)$, in the {\it pQCD 2 gluon model} $(b.)$ and in the
SCI approach $(c.)$.
In $(a.)$ a typical tree level ${\cal O}(\alphas^2)$ diagram is shown.
}\label{diagram}
\end{figure}
 
This approach has its origin in the Ingelman-Schlein model, which says,
that diffraction is mediated by the pomeron \PO\  which behaves
like a particle and consists of partons, which can be described
by parton density functions (pdf)  evolved with the  
DGLAP evolution equations.
Hard scattering processes, like jet - or heavy quark production occur
according to the same hard scattering matrix elements as
known from standard $ep$ and $p\bar{p}$ scattering. In general,
it is enough to define diffractive parton
density functions $f^D(\xi,\xpom,t;Q^2)$, 
without making any assumption about the pomeron. 
\par
Collins showed, that diffractive processes, 
induced by scattering of a (direct) photon, 
can be factorized into these diffractive pdf's $f^D(\xi,\xpom,t;Q^2)$ and hard
scattering coefficient functions, as known from standard QCD processes:
\begin{equation}
d \sigma = \sum_i \int d \xi f^D(\xi,\xpom,t;Q^2) d\hat{\sigma}_i
\end{equation}
up to corrections that are power suppressed in $Q$, with $\xi$ being the
fractional momentum of the struck parton relative to the pomeron, 
$\xi = x / \xpom$. 
\par
Important is the statement, that the factorization theorem is 
valid {\it only} for direct photon and leading $Q^2$ contributions. 
Nothing is said about the relative
size of the non-leading contributions, and whether they are larger or smaller 
compared to inclusive processes. This factorization
theorem is expected not to be valid in the case of $p\bar{p}$ or resolved photon
interactions.
\par
The diffractive parton densities $f^D(\xi,\xpom,t;Q^2)$ are determined by 
LO or NLO DGLAP
fits to the diffractive structure function $F_2^{D(3)}(\xi,\xpom;Q^2)$. 
With NLO diffractive parton densities
also NLO order programs might be used for
diffractive di-jet production as well as for diffractive charm. However, it is
necessary to treat the {\it pomeron remnant} properly, as the details affect the
formation of the rapidity gap, which often forms the basis for the calculation
of \xpom\  (see Fig.~\ref{diagram}). 
In the case of charm production, additionally one has to treat
properly the formation of one or two $D^*$-mesons and also
of bound $c\bar{c}$ states, which are not covered by
applying naively the fragmentation functions. 

\subsubsection{Two-gluon approach} 

Diffractive final states, which consist of hard jets or heavy quarks
(without soft remnants) can be fully perturbatively calculated because 
the hard scale which allows the use of pQCD is provided by the
large transverse momenta of the jets, and the Pomeron exchange is modeled
by the un-integrated gluon density of the proton.
Recently, full perturbative calculations were
done for $e p \to e'\; q\bar{q} \;p$ and $e p \to e'\; q\bar{q}g \;p$, where $q$
can be a light or heavy quark~\cite{kyrieleis}
(see Fig.~\ref{diagram}). This approach provides a natural connection to
non-diffractive scattering as the same un-integrated gluon density is used,
without any new parameters.
\par
These perturbative calculations have been implemented in the \RAPGAP\ Monte 
Carlo
generator, supplemented with final state parton showers and full hadronization.
Recently, also different approaches to parameterize the un-integrated gluon 
density
were published, which are applicable also to the region $\kt \to 0$. This region
is important in the calculations, as the un-integrated gluon density appears
inside an integral over an internal loop.   

\subsubsection{Soft Color Interaction model}

The soft color interaction model (SCI)
and the generalized area law model (GAL) 
were developed under the assumption that soft
color exchanges give variations in the topology of the color strings such that
different final states could emerge after hadronization, e.g. with and without
rapidity gaps or leading protons. 
The same hard scattering processes as in
non-diffractive scattering are used, with the same non-diffractive parton
density functions.
In the SCI model an explicit mechanism is applied to
exchange color between the emerging partons, in the GAL model color is exchanged
between the strings. The probability for the soft color exchange to happen, is a
free parameter which has to be determined from experiment. 
\par
The SCI and GAL models are successful in describing $F_2^{D(3)}$ as well as 
diffractive jet production at the Tevatron~\cite{nicusor}.
\par
However, it has been argued ~\cite{lonnblad}, that using the standard 
DGLAP pdfs of the proton together with DGLAP parton showers 
to describe diffraction might be problematic.
Especially in the forward
region of $\eta \sim 3 $, which is often used to select diffractive events, and
which corresponds to $\xpom \leq 0.05$, the standard DGLAP approach already
fails to describe non-diffractive processes, like forward jet production. 
Diffractive scattering could  be sensitive to small
$x$ effects in the parton distribution functions, as predicted by BFKL or CCFM.

\subsection{What is new ?}

It is interesting to note, that all three approaches and models to describe and
understand diffractive scattering have been basically presented already in 1995,
where the first detailed measurements of the diffractive structure function
$F_2^{D(3)}$ has been presented. 
\par
Besides the measurements of vectormeson production and
 $F_2^D$~\cite{capua,laycock} with the pdf
fits~\cite{schilling}, new measurements of charm production have been
performed~\cite{vlasov} and compared to predictions using diffractive parton
densities.  
\par
Now having new and more precise measurements on charm and jets available, tests
of the factorization theorem can be performed as well as more detailed
investigations of the structure of the diffractive hadronic final state. The
discrepancy in description of di-jet and charm is suggestive for mechanisms of
factorization breaking and deviations from the DGLAP picture. As discussed
below, the 2-gluon picture could provide an explanation of these effects. 
However one has to
keep in mind, that the errors both on the data and also on the diffractive pdfs
are still too large to draw final conclusions.

\subsection{Factorization tests}

One of the most important issues in diffraction is the universality of
diffractive parton density functions:
\begin{itemize}
\item To which extend is the factorization theorem in deep inelastic 
diffraction
satisfied ? Is it similar to non-diffraction or are there differences?
\item Are pdf's obtained in $ep$ scattering applicable to $p\bar{p}$ 
scattering ?
\end{itemize}

To answer these questions, the diffractive parton densities have to be 
determined precisely. Then these pdf's can be used in other processes to compare
predicted cross sections with the measured ones.

\subsubsection{Factorization tests at HERA}

The diffractive parton densities are used in the {\it resolved pomeron} approach
to calculate  inclusive diffractive
cross sections as well as the cross sections for jet and charm production. 
For practical reasons it is important to perform all calculations within the
same framework and program, as different extrapolations from the measured to the
hadron or parton level cross section might introduce additional uncertainties.
Such extrapolations can only be performed properly by the use of full hadron
level Monte Carlo generators, 
and different approaches within the same framework
help to estimate the model uncertainties.
\par
In general the MC predictions obtained in the {\it resolved pomeron} approach
agree very well with the measurements, 
as shown in detail in~\cite{thompson}. Using the (old) H1 fit2 as a 
parameterization of the diffractive parton densities, di-jet production in 
diffractive DIS~\cite{h1_diff_dijets} is well described, if resolved virtual
photon contributions are included to simulate the {\cal O($\alpha_s^2$) } 
(NLO) corrections. 
However, using the same parameter set (and the same Monte
Carlo event generator) the predicted cross section for 
charm production~\cite{h1_diff_charm} overshoots the data.
\begin{table}
\begin{center}
\begin{tabular}{ |c|c|c| }
\hline
     & H1 fit2                                 & New H1 fit \\ \hline
 H1  & $0.67 \pm 0.15 \pm 0.15 $                 &
  $1.23 \pm 0.27 \pm 0.28 $\\ \hline
 ZEUS& $0.63 \pm 0.07 ^{+0.04}_{-0.08} \pm 0.03$ &
     $1.31 \pm 0.12  ^{+0.08}_{-0.16} \pm 0.05$ \\ \hline
\end{tabular}
\end{center}
\caption{The ratio $R^{D^*} =\sigma(\mbox{data})/\sigma(\mbox{MC})$ for
diffractive charm production of H1 and ZEUS~\protect\cite{vlasov} 
obtained from the H1 fit2 and the
new H1 fit~\protect\cite{schilling} presented at this conference.
\label{ratio}}
\end{table}
A new and precise measurement of
charm production in diffractive DIS, performed by ZEUS, has been presented 
at this conference~\cite{vlasov}.
Using the H1 fit2 the Monte Carlo predictions overshoot the data significantly,
whereas with the new
set of diffractive pdf's, diffractive charm production in 
deep inelastic scattering is reasonably well described (Tab.~\ref{ratio}).
\begin{figure}
{\scalebox{0.8}{\includegraphics*{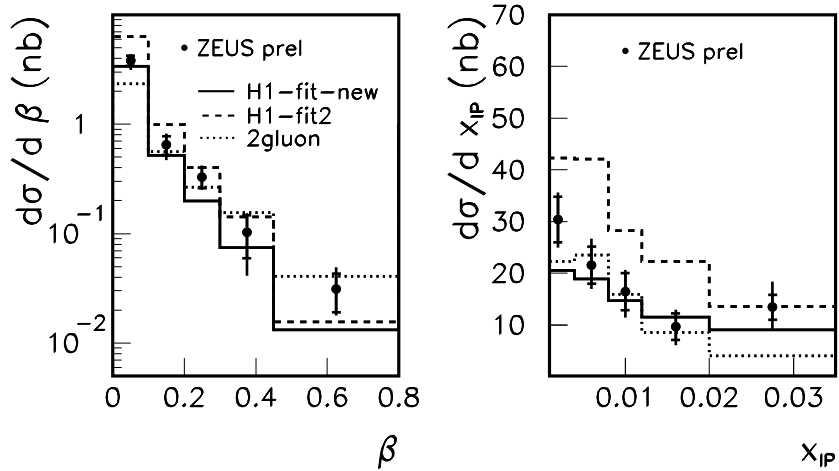}}}
\caption{The cross section of charm production in deep inelastic diffraction as
measured by ZEUS\protect\cite{vlasov} compared to the prediction from \RAPGAP\
using the old and the new diffractive pdf's as obtained by
H1~\protect\cite{schilling-private}.
Also shown is the prediction of the perturbative 2-gluon approach with the
unintegrated gluon from the saturation model.
}
\label{charm_dis}
\end{figure}
In Fig.~\ref{charm_dis} we show the comparison of the ZEUS diffractive charm
measurement with the predictions from the resolved pomeron model using the old
and the new diffractive pdf's obtained by H1 \cite{schilling-private}.
However, applying the same diffractive pdf's to diffractive di-jet
production, the predicted cross section is smaller  
compared to the H1 measurement~\cite{h1_diff_dijets}. 
It is suggestive that even with the new pdf set a  difference is observed 
in the  description of diffractive final state processes. 

\subsubsection{Factorization breaking at HERA?}

The difference in description of diffractive di-jet and charm measurements
within the {\it resolved pomeron approach} might be a possible sign for factorization
breaking, as the corresponding quantities are well described in non-diffractive
$ep$-scattering. 
A possible mechanism to
explain factorization breaking between diffractive charm and di-jet production
is provided by the perturbative calculation of 2 gluon exchange processes: with
a reasonable choice of the un-integrated gluon density, both charm and di-jet
production can be described using the same set of parameters. This becomes
understandable since so called  {\it non-$\kt$-ordered}  gluon
emissions contribute differently to the di-jet and charm sample. 
In the kinematic range of the analysis, 
the perturbative 2-gluon calculation \cite{kyrieleis} 
(as implemented in \RAPGAP ) 
yields much more  events  having a gluon with transverse
momentum larger than those of the quarks: $\sim 30 $\% for diffractive di-jets 
and 
$\sim 15 - 17$ \% for diffractive charm, compared to 2 - 3\% obtained in the
resolved pomeron mode, depending on the choice of the factorization
scale $\mu^2$. 
\par
The calculation has been previously compared with the measurement of
diffractive di jet production, and reasonable agreement has been found in the
region of $\xpom < 0.01$, if the transverse momenta of all partons are
required to be larger than $\kt \gap 1.5 $ GeV. Especially the gluon in the 
$ q\bar{q}g $ process is found to be hard, in contrast to the expectation in the
resolved pomeron model, where this gluon should appear as a soft pomeron
remnant. As presented in this workshop, the perturbative calculation has also
been compared with the measurement of diffractive charm production
\cite{kyrieleis,vlasov}. For both the H1 and ZEUS measurement reasonable
agreement has been found using the same cutoff of $\kt \gap 1.5 $ GeV
for the gluon transverse momentum.
A comparison with the new charm measurement of ZEUS is shown in
Fig.~\ref{charm_dis}. 
\par
The above considerations show, that a detailed simulation of the hadronic final
state in diffraction is inevitable when going beyond total cross section
calculations.

\subsubsection{Factorization tests: HERA versus Tevatron}

As already mentioned, the measurement of diffractive di-jet production at the 
Tevatron 
compared to the prediction using diffractive pdf's obtained from HERA
can be used directly to test the factorization hypothesis.
\par
 CDF~\cite{terashi} has measured the measured diffractive
structure function in the kinematic region $|t| <1$ GeV$^2$, $0.035 < \xi
< 0.095$, and $E_T^{jet1,2}> 7$~GeV. The comparison between the HERA
extrapolation to the CDF kinematical domain 
 shows a discrepancy in normalization of a factor 7 to 8
between both experiments, see Fig.(\ref{compare}). However, contrary to previous 
results~\cite{terashi} based on
old H1 diffractive pdfs, the shape of both distributions seems to be quite
close~\cite{terashi,schilling}. It seems that these results could be interpreted in term
of a gap survival probability which does not seems to be kinematically
dependent. 
\par
However,
the observation of non-factorization at $p\bar{p}$ does not come as
a surprise as already in the proof of the factorization 
theorem it has been
noted that this only applies to direct photon interactions.

\subsubsection{Global fits of HERA and Tevatron data}

An attempt to fit together HERA and Tevatron data has been performed in Ref.
\cite{schoeffel}. It was found that it is impossible to fit both sets of data
together even by letting free the normalization between both experiments.
Unfortunately, this study has been performed using the old published 1994 H1
diffractive structure function data, and it would be worth to redo it using the
preliminary 1997 data, since they lead to different quark and gluon densities in
the pomeron. 
A gluon density has also been extracted directly from CDF data and
found to be more proton-like, with a softer value of the intercept, 
than the diffractive gluon density at HERA \cite{schoeffel}.
\par
It is interesting to note, that in contrast to single diffraction mention above,
the 
mass fraction measured in double diffractive events by the CDF collaboration is
compatible with the HERA gluon density and not with a proton-like structure
function. It seems that double diffraction at Tevatron is harder than single
diffraction. 
More data from Tevatron are needed to do more
studies related to the dijet mass fraction, which will happen very soon
with the start of Run II.

\section{Diffraction and QCD Theory}
There are several  connections with basic problems in QCD raised by
experimental 
and phenomenological studies on diffraction, but they are kind of subtle.
Indeed, ``hard diffraction'' being a superposition of a short space-time
interaction at weak coupling (at the hard $\gamma^*,$ dijet, Higgs,...vertex) 
with a typically soft process 
at   strong coupling (at the soft vertex with the 
ingoing proton or antiproton), it is highly 
non trivial to formulate it in fundamental terms. On the other side of the same 
coin, hard diffraction may represent a new route to theoretical progress in the 
interface 
between soft and hard QCD, and thus for a more complete, and presently yet
unknown, formulation of the theory.
 
\subsection{Small-x: questions and  progresses}
One important theoretical aspect where diffraction can teach us something is the 
never-ceasing discussion on the physical relevance (or not) of the BFKL type of 
evolution equation. Indeed, diffractive processes represent a good opportunity of 
so-called ``one-scale'' configurations, where the ``BFKL logs'' are favored 
w.r.t. the usual DGLAP evolution. This is well illustrated in the study of 
vector meson production. Deeply virtual Compton scattering adds a new aspect of 
the problem by the comparison of QCD dipole models (neighboring BFKL 
evolution) with Generalized pdf's  (extending DGLAP evolution).

\subsubsection{ Is BFKL ``seen''? : Vector meson production}

Diffractive vector meson production at large momentum transfer $t$ accompanied 
by proton dissociation is a good laboratory for studying resummation of
perturbative QCD effects in color singlet exchange reactions. Indeed, this is 
a 
one-hard-scale process which is expected to be described by the non-forward 
BFKL 
equations. However, the usual leading $\log 1/x_{Bj} \sim Y$ approximation (Y 
is 
the rapidity gap interval), corresponding only to the leading ``conformal 
spin'' 
$n$ of the expansion of the BFKL equation's solution, fails  to reproduce the 
data. In \cite{enberg}, it is shown that the inclusion of the whole $n$ 
expansion gives a good description of data, see Fig.~\ref{Feyn}, and also a 
reasonable 
one 
of spin density matrix elements. It is not clear, however, how the 
next-to-leading  $\log x_{Bj}$ contribution, which is known to be large at 
least at  $t=0,$ 
would modify the conclusions. An argument is that the simplest Feynman 
graph in Fig.~\ref{Feyn}, \ie the 2-gluon exchange, includes already an all-$n$ 
expansion.
More work is deserved in this interesting direction, including large $t$ 
processes at Tevatron. A puzzle to be solved is the value of the strong 
coupling 
constant in the cross-section pre-factors, which ought to be {\it non running} 
for phenomenological applications, while common wisdom would expect an 
improvement with usual running behavior. 
\bfig
\bc
\epsfig{file=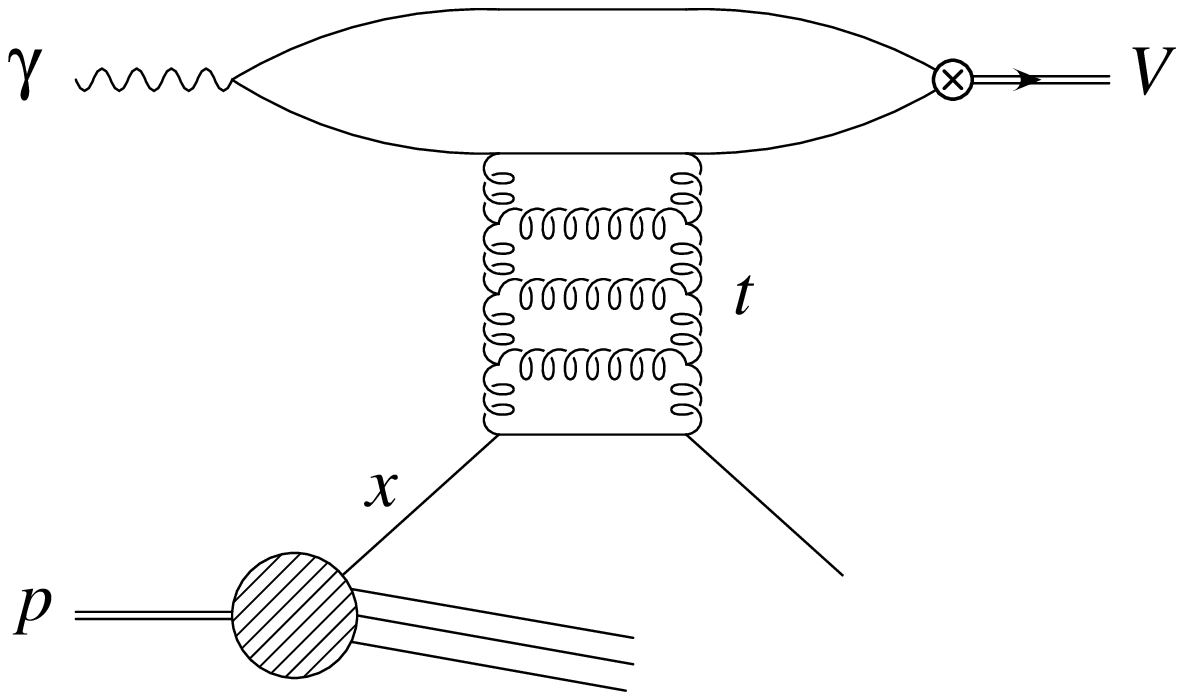, width=0.49\columnwidth,
        bbllx=173, bblly=286,bburx=510, bbury=517}
\epsfig{file=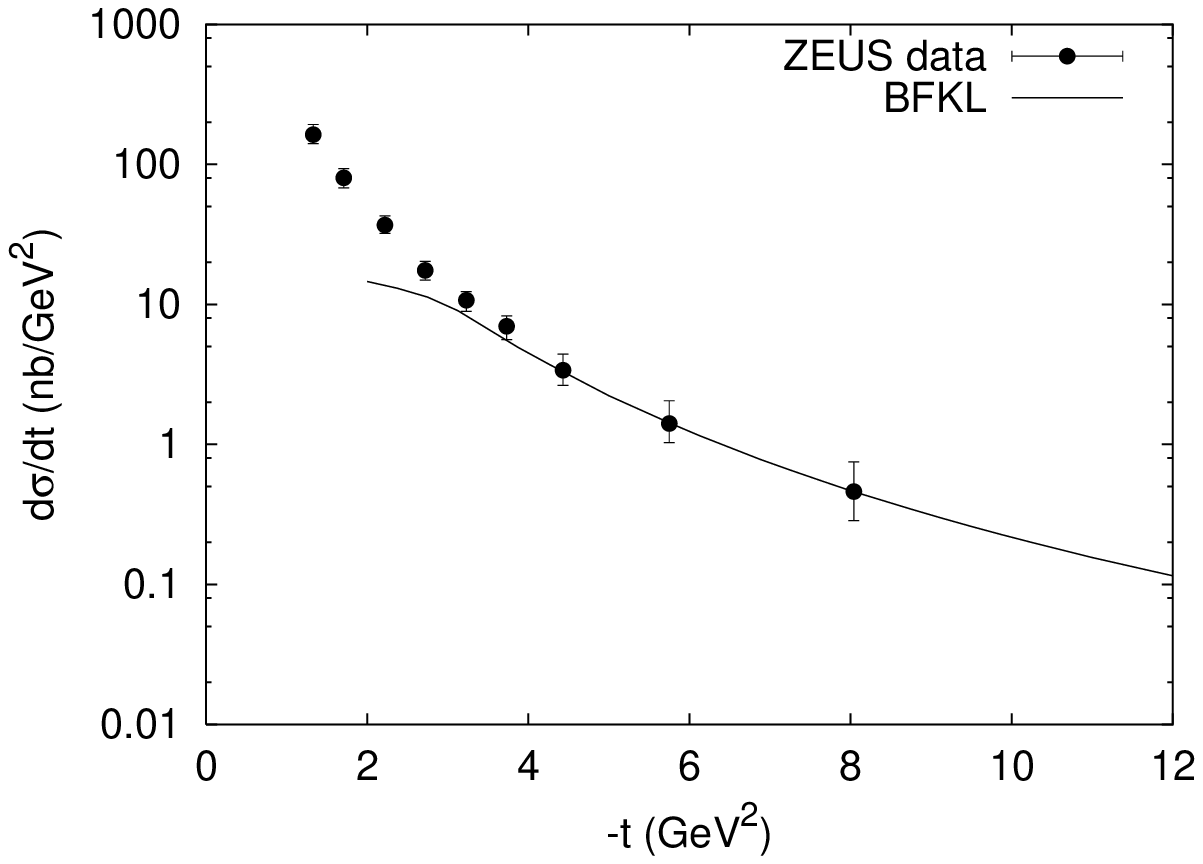, width=0.5\columnwidth}
\ec
\caption{Feynman graphs and differential cross-section for 
the process $\gamma p\to\rho X$ at large $t$ (from \cite{enberg}).}
\label{Feyn}
\label{rhosigma}
\efig

At smaller momentum transfer and in total $\gamma ^*p\to V\ p$ cross sections, 
the hard scale is provided by $Q^2.$ In BFKL physics, the coupling to the 
proton 
is {\it via} the un-integrated gluon distribution, where track is kept of the 
transverse momentum of the gluon. This ``$k_T$-factorization'' property allows 
to relate  the studied processes to the proton structure function in the BFKL 
formalism. In \cite{ivanov},  both non-perturbative and perturbative QCD 
contributions to the un-integrated gluon distribution have been considered, 
where soft and hard components are strongly separated. The comparison with 
cross-sections shows discrepancies while ratios, \eg $\sigma (\phi)/\sigma 
(\rho)$ are in better shape. One remark is that progress on the various 
determinations, models and theoretical properties of un-integrated gluon 
distributions are highly required in this field.

\subsubsection{Dipoles or Generalized pdf's? : $\gamma ^*\ p \to \gamma\ p$}

\begin{figure}[htbp]
   \includegraphics[width=5.5cm,height=4.5cm]{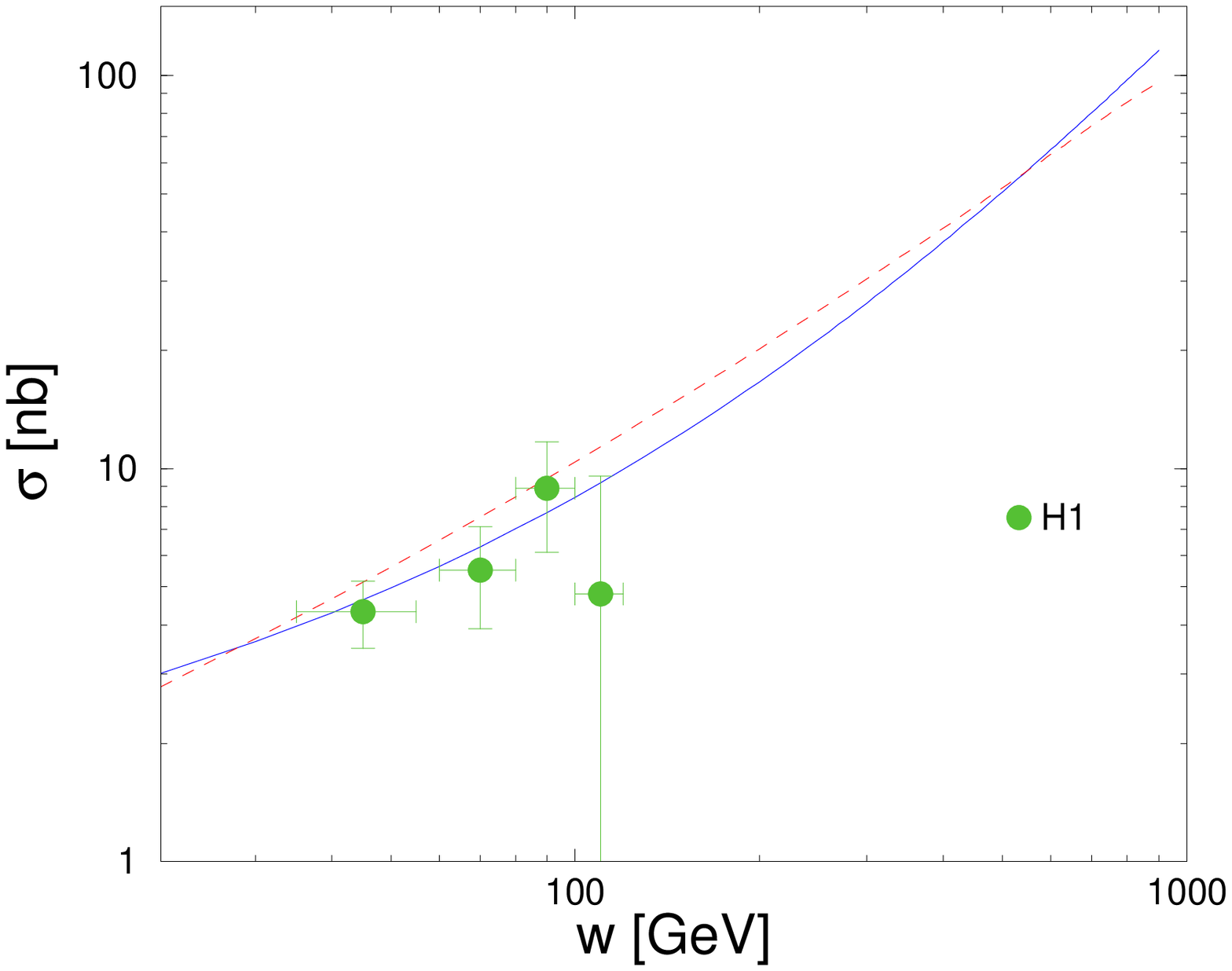}
  \hspace{1cm} 
\includegraphics[width=5.5cm,height=4.5cm]{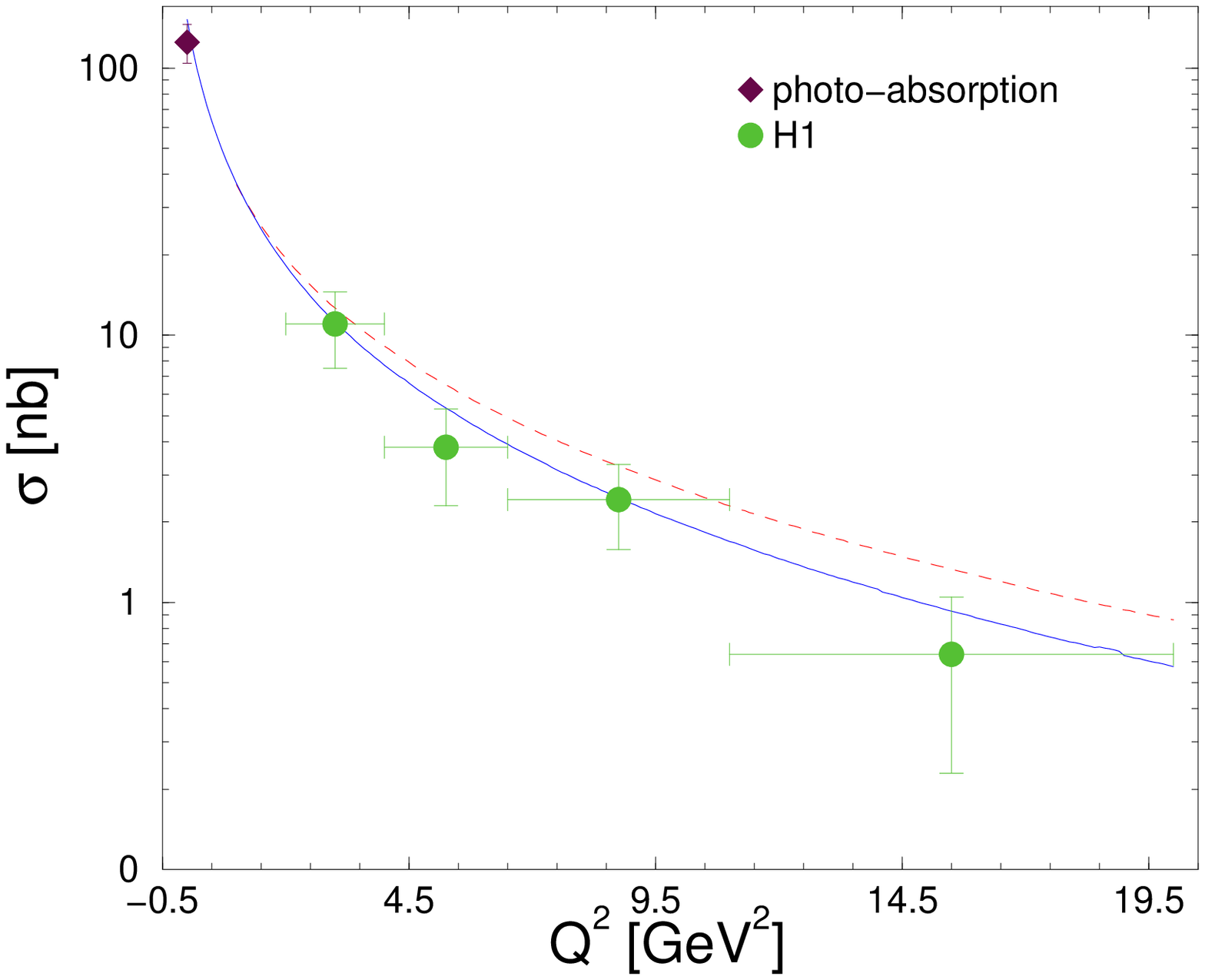}
  \caption{{\it The energy (at $Q^2=4.5~\mbox{GeV}^2$) dependence and 
$Q^2$-dependence (at $W = 75$~GeV) of the photon level DVCS cross-section, FKS 
(solid line) and MFGS (dashed line) (from \cite{sandapen}).}}
    \label{totsig.eps}
\end{figure}

Color dipole models accommodate rather easily the description of DVCS ($\gamma 
^*\ p \to \gamma\ p$) processes. In \cite{sandapen}, for instance, two 
different 
models for the dipole-proton cross-sections, one with saturation included 
(MFGS) 
and one without (FKS) give  good predictions for DVCS data, see 
Fig.~\ref{totsig.eps}, based on 
non-perturbative inputs for {\it other} processes. On the other hand, the 
approaches based on NLO QCD evolution equations \cite{mc}, which are more 
solid 
on a theoretical point of view, seem however to depend much \cite{freund} on 
the 
non-perturbative input for the {\it same} process, even if good candidates have 
been found. On a theoretical ground, it will be worthwhile to understand better 
the connections between the two approaches, the former being ``$s$-channel'' and 
the 
latter ``$t$-channel'' oriented. DVCS can then be a good laboratory for the much 
discussed comparison of ``$s$-channel'' and  ``$t$-channel'' models.
The comparison between elastic and inelastic diffraction has been wisely 
advocated to be important for the validity of dipole models. A variant of the 
model, motivated by  generalized vector dominance \cite{schild}, is shown to 
lead to a specific geometrical scaling prediction for elastic diffraction and a 
good prediction for the longitudinal (high $\beta$) part of
inelastic diffraction, while the transverse 
photon polarization requires higher spin components of the hadronic states. 
More 
work disentangling the various versions of dipole models is certainly deserved 
in the near future. Another type of dipole model has been advocated \cite 
{uter}, which relies on the introduction of instantons to describe the soft QCD 
regime of the dipole-proton diffractive coupling. All this emphasizes  the 
importance of a good understanding of the dipole-proton cross-section both in 
the perturbative and non-perturbative regimes.

A relevant point for the discussion is the amount of higher-twist with respect 
of leading twist contributions in diffractive processes. It is known that the 
``$s$-channel'' naturally contain larger   higher-twist components than 
``$t$-channel'' ones. Nuclear shadowing has been suggested \cite{mc} to give 
experimental separation between these two options by looking to nuclear gluon 
pdf's through the analysis of $F_{2A}/AF_{2N}$ at very small $x_{Bj}$ when 
possible. Note also~\cite{gol} the interest of double spin asymmetries in 
diffractive $Q\bar Q$ production.

\subsection{Smaller-x: Saturation, myth or reality?}

The question of {\it saturation} has been a major problem discussed in the 
conference. For
$x_{Bj}\to 0,$ the growing number of gluons of fixed size $1/Q$ in the 
wave-function of the proton  becomes high enough that new multi-parton 
interaction occurs. These effects  modify the evolution equations by the 
addition of non-linear terms and even may lead to a new phase of QCD.

\subsubsection{Has Saturation already been seen at HERA?}

Despite the existence of 
an elegant and inspiring model of saturation \cite{GBW1}, there is not yet any 
compelling 
evidence of saturation in inclusive structure function data. Hence it is of 
primary importance to look for model-independent investigations on saturation 
 with less inclusive data at HERA.
 
\begin{figure}[ht]
\begin{center}
\epsfig{file=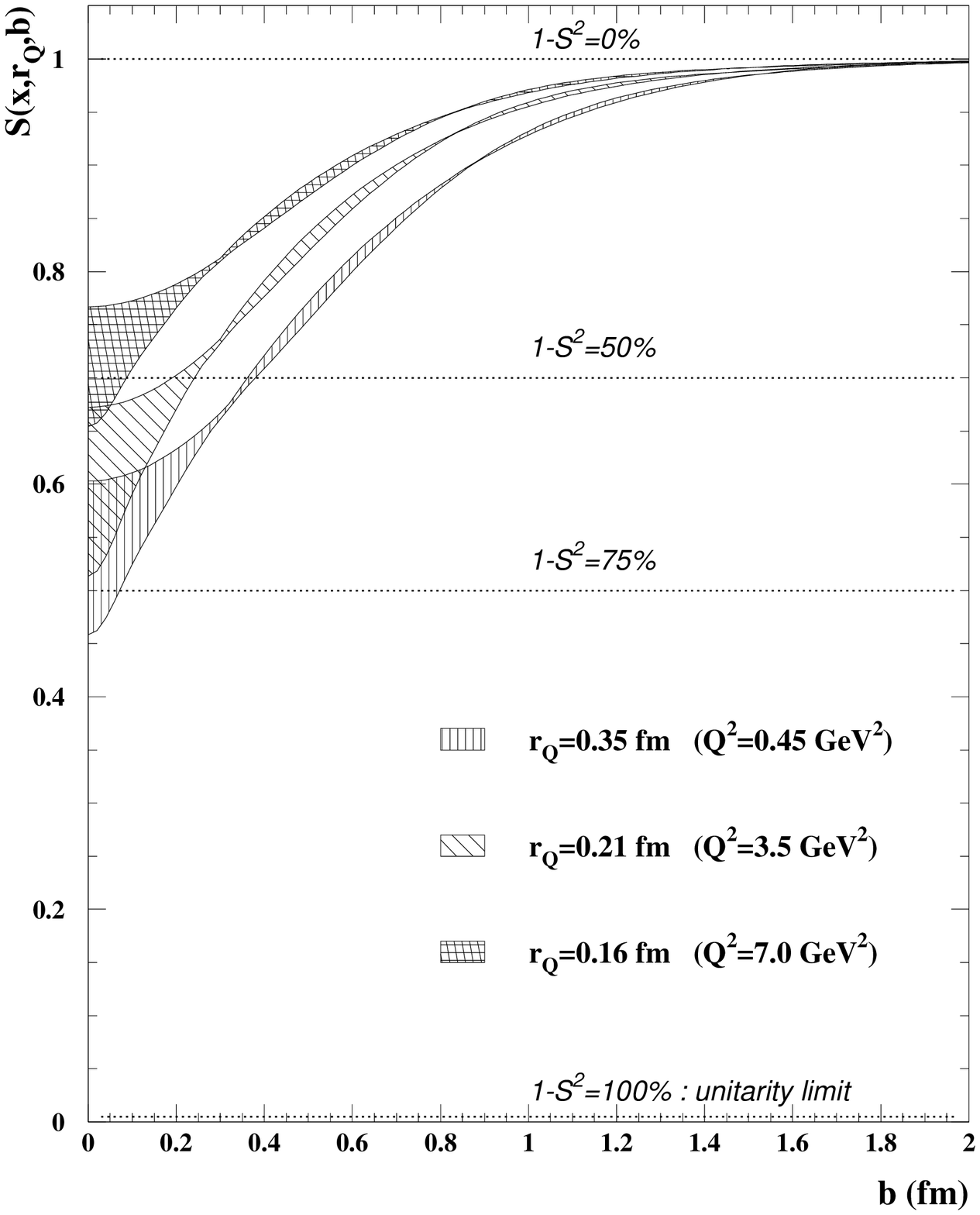,width=8cm}
\end{center}
\caption{{\it $S$-matrix as a function of the impact parameter
for $x_{Bj}\!\sim\!5\cdot 10^{-4}$ and $Q^2=0.45,\ 3.5,\ 7\ \mbox{GeV}^2$ (from 
\cite {munier}).
The width of the bands represents the uncertainty due
to the lack of experimental data for $t>0.6\ \mbox{GeV}^2$. 
It is obtained by extrapolating the cross section
with functions of $t$ 
with behavior between $t^{-3}$ and $e^{-\lambda t}$.
}}
\label{1}
\end{figure}

This has been 
proposed in \cite{munier} using diffractive $\rho$ production cross-section on 
a 
proton as a function of $t.$ The 
method is to extract the $S$-Matrix elements $
S(x_{Bj},r_Q,b),$ where $r_Q $  is the mean dipole size probed by the $\rho$ 
wave-function and $b$ is the impact parameter of the reaction. Interestingly 
enough, see Fig.~\ref{1}, perturbative saturation (likely to be reached when   
the 
interaction
probability $1\!-\!S^2$ is large while $1/Q$ remains small) seems to be 
relevant
at small $b.$  This might be interpreted as a first experimental evidence for
saturation~\cite{munier}.
Complementary  confirmation is highly deserved for this first 
experimental evidence for saturation at small $b$ at HERA.
  
The linear DGLAP evolution equations are known to work well for inclusive 
structure function data. The saturation models incorporating non-linear 
contributions have thus to match  this  constraint. This uneasy problem has 
found in the conference two developments. The original saturation model 
\cite{GBW1} has been modified to take into account the linear evolution at 
small dipole size \cite{golec-dis2002}. 
It indeed leads to a better description of $F_2\sim 
x_{Bj}^{-\lambda(Q^2)},$ see \cite{struc}. Another proposal \cite{kwiecinski} 
is 
to match the linear and the non-linear behavior on the saturation critical 
line $Q=Q_S(x_{Bj}).$
However, these proposals lead to modifications of the 
simple structure of the original model, \eg the geometrical scaling behavior 
\cite{kwiecinski}, which have to be confirmed/disproved by further 
phenomenological and theoretical investigations. 

\subsubsection{The Color Glass Condensate: a new QCD phase?} 

\bfig
\bc
\epsfig{file=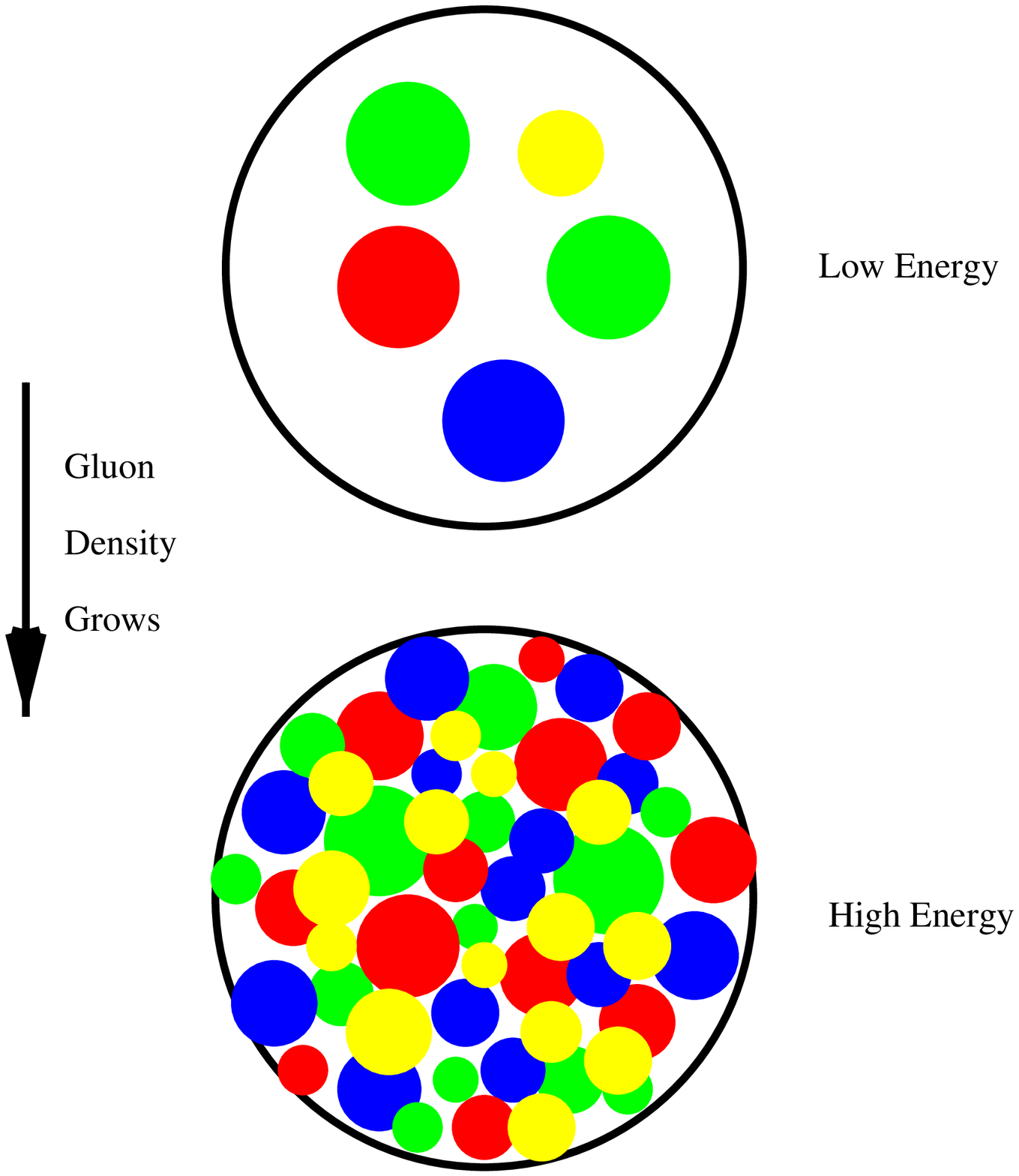, width=0.45\columnwidth}\hspace {2cm}
\epsfig{file=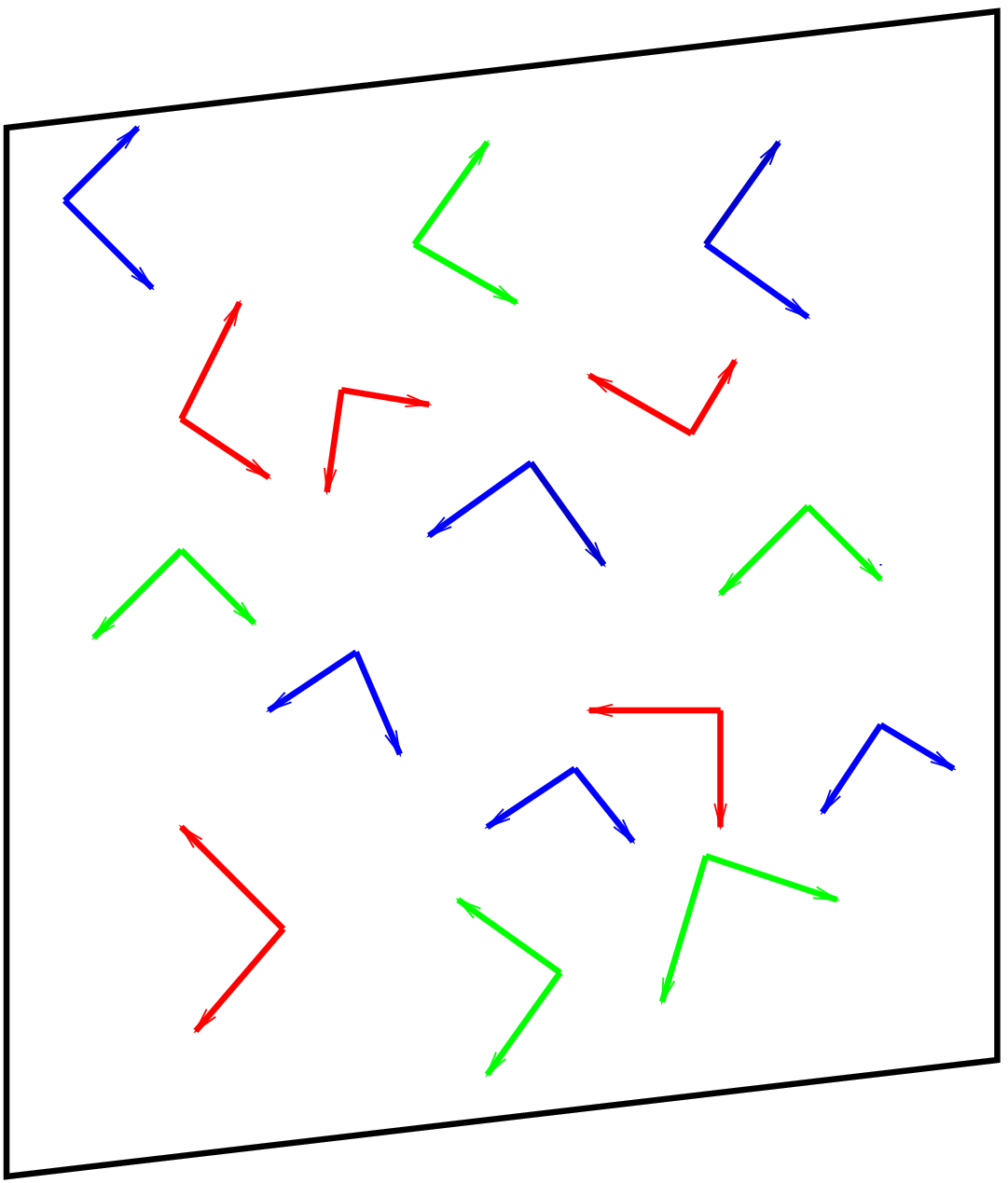, width=0.25\columnwidth}
\ec
\caption{{\it Saturation at small $x_{Bj}$ and the ``color glass condensate'' 
(from \cite {mcl}). 
Arrows schematically describe the classical QCD  fields.}}
\label{saturation}
\efig
As we have seen, the evolution of the gluon distribution in a proton when 
$x_{Bj}\to 0$ is expected to lead to an over-density  in the geometrical phase 
space of the proton, see Fig.~\ref{saturation}.
 The question arises then how to describe the new 
state with high density and weak coupling in QCD. A  proposal has received the 
name ``color glass condensate'' \cite {mcl}. The main idea is to consider that 
the abundance of gluons legitimates a classical colored field approximation  
(see Fig.~\ref{saturation}). 
For each  density 
increment due to the evolution to smaller $x,$ new gluonic states are generated  
in the background of the classical 
field 
configuration. This  gives rise to a non-linear 
evolution equation which can be interpreted as a high-density  modification of  
the linear BFKL equation. 
The density increment is attributed either to an increase of nuclear size ({\it 
i.e.} a change of  initial conditions) or 
to evolution towards small $x_{Bj}$ ({\it i.e.} a change in the  quantum 
evolution). This 
structure recently led to many applications such like geometrical scaling, saturation in 
hadron-hadron and heavy ion collisions, etc....

It would be interesting to disentangle, either theoretically or 
phenomenologically, the actual quantum evolution aspects from those coming from 
initial 
conditions.
Another pending question is the relation, which has been recently 
advocated \cite{mcl} (but also criticized \cite{urs}), between saturation and 
the 
Froissart bound. The Froissart bound comes from unitarity, analyticity and the 
existence of a ``mass gap'' in the theory, namely between zero and the pion as the lowest mass 
state. This last requirement, related to confinement,  is not 
apparent in the saturation picture. 

\subsection{Smallest-x: the ``Regge'' mystery in QCD} 

Since a long time ``Regge behavior'' has been  a great challenge for our 
understanding of high energy reactions. It mainly consists of $2 \to 2$ 
particle 
amplitudes behaving as $A^{el} \sim e^{\alpha_{\PO}(t)\ Y}$ for vacuum 
exchange reactions (also named {\it Pomeron}) and 
$A^{inel} \sim e^{\alpha_{\RE}(t)\ Y}$ 
for various non vacuum quantum number exchanges (also named {\it 
Reggeons}). For soft hadronic reactions, the ``Regge trajectories'' 
$\alpha_{\PO,\RE}(t)\sim \alpha_{\PO,\RE}(0) + \alpha'_{\PO,\RE}\ t$  are 
thus found to be phenomenologically linear, with universal intercepts 
$\alpha_{\PO,\RE}(0)$ and slopes $\alpha'_{\PO,\RE}$ for a given set of 
exchanged quantum numbers. Since long, there are attempts to derive the Regge 
behavior from resummation of the perturbative QCD expansion at large $Y,$ 
while, recently, new tools for non-perturbative QCD estimates have been 
experienced. Note that the ``Regge behavior'' has become even more 
mysterious by the 
experimental evidence for the $Q^2$-dependence of effective intercepts and slopes for 
hard $2 \to 2$ processes, such as $\gamma^*\ p$ cross-sections.

\subsubsection{The origin of ``Regge behavior'' in  pQCD} 

\begin{figure}[ht]
\vspace*{3mm}
\centerline{{\epsfysize 3.2cm \epsfbox{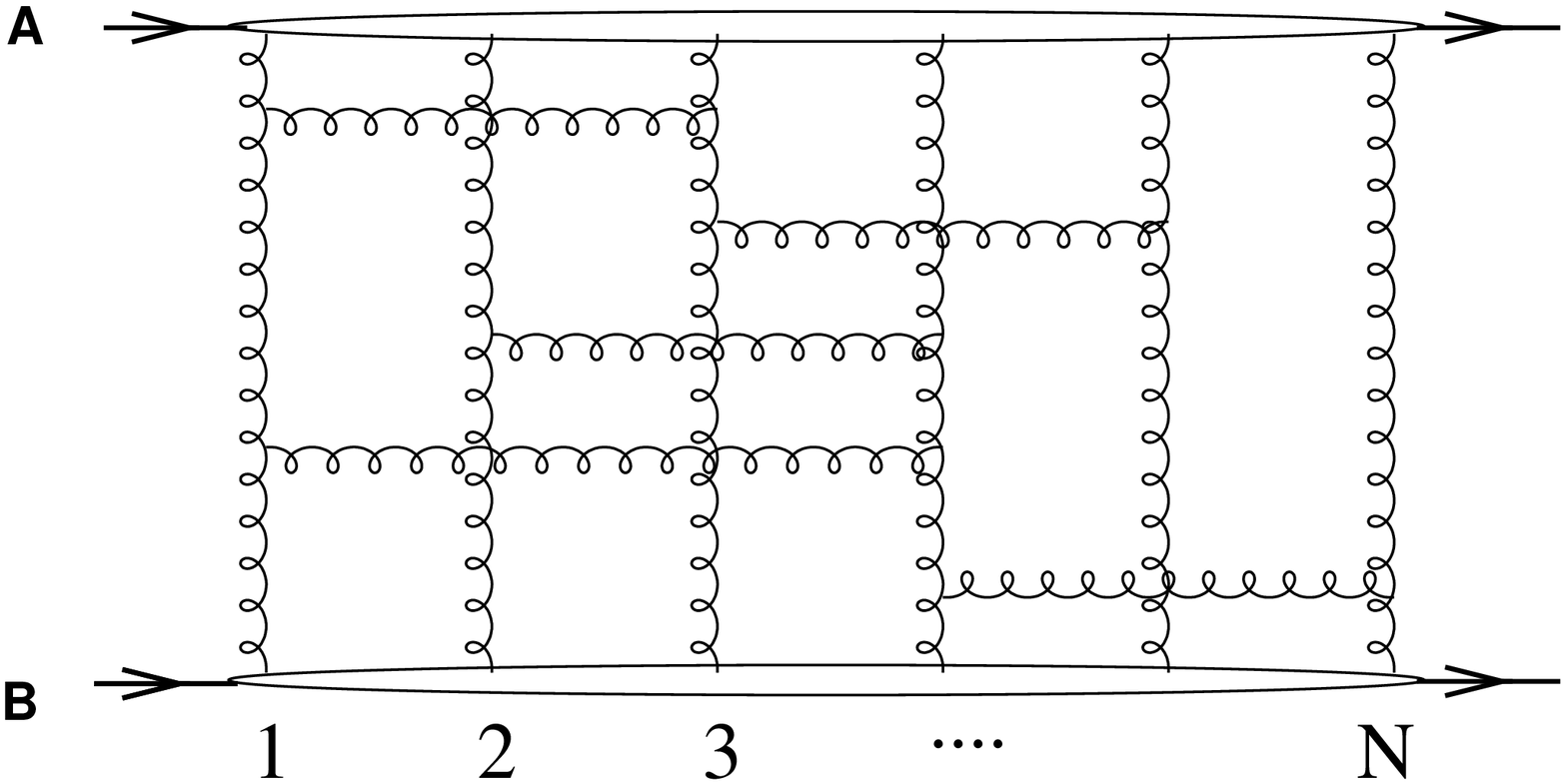}}
\quad{\epsfysize 4.0cm \epsfbox{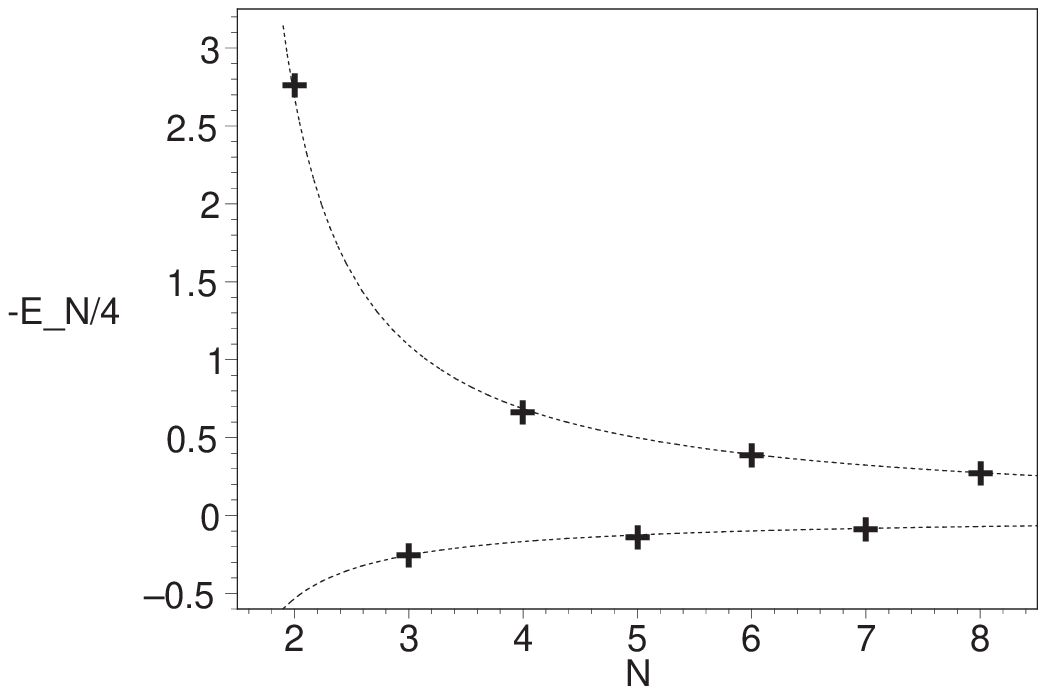}}}
\caption[]{{\it Feynman graphs and  dependence of the intercept 
$\epsilon_N$, on the number of
gluon ladder up-rights $N$ (from\cite {kot}). }}
\label{en}
\end{figure}
 
On the pQCD side of the question, the challenge is to sum Feynman diagrams 
contributing to the large $Y$ behavior of amplitudes. Generalizing those of 
Fig.~\ref{Feyn}, 
with $2$ gluon up-rights corresponding to the BFKL equation, one needs to 
resum those of Fig.~\ref{en} 
with $N$ gluon up-rights (and gluon	rungs connecting any 
two 
of them), which is a formidable task. In fact, a solution for intercepts has 
only now been found \cite{kot} and reported at the conference.  The 
mathematical 
breakthrough was the recent  formulation of the problem as an integrable 
quantum 
spin chain with $N$ sites and ground energies $\epsilon_N.$ However from 
``integrable'' to ``integrated'' required a lot of work. As a result, one 
obtains a series of intercepts $\alpha_{N}(0) \equiv \frac {g^2 N_c}{4\pi}\ 
\epsilon_N$ which, for an even number of gluon up-rights (see 
Fig.~\ref{en}),
 contribute to 
the {\it Pomeron} sector, while for an odd number of gluon up-rights, contribute 
to 
the elusive {\it Odderon} which has not been yet found experimentally. 
Interestingly enough, all Pomeron intercepts are greater than $1$ (growing 
contributions to total cross-sections), while all Odderon ones are less than 
$1.$ Note however another solution \cite{lipatov} with Odderon intercepts equal 
to one. These theoretical investigations should lead to a deeper understanding 
of the connection of the Regge behavior with QCD. For instance the couplings of 
external sources, and the dependence on the hard scale (which is fixed in 
actual 
calculations) are both problems to be addressed soon. Summing non planar 
diagrams which are known to contribute is yet another deep challenge.
\begin{figure}[htbp]
\begin{center}
\includegraphics[width=6cm]{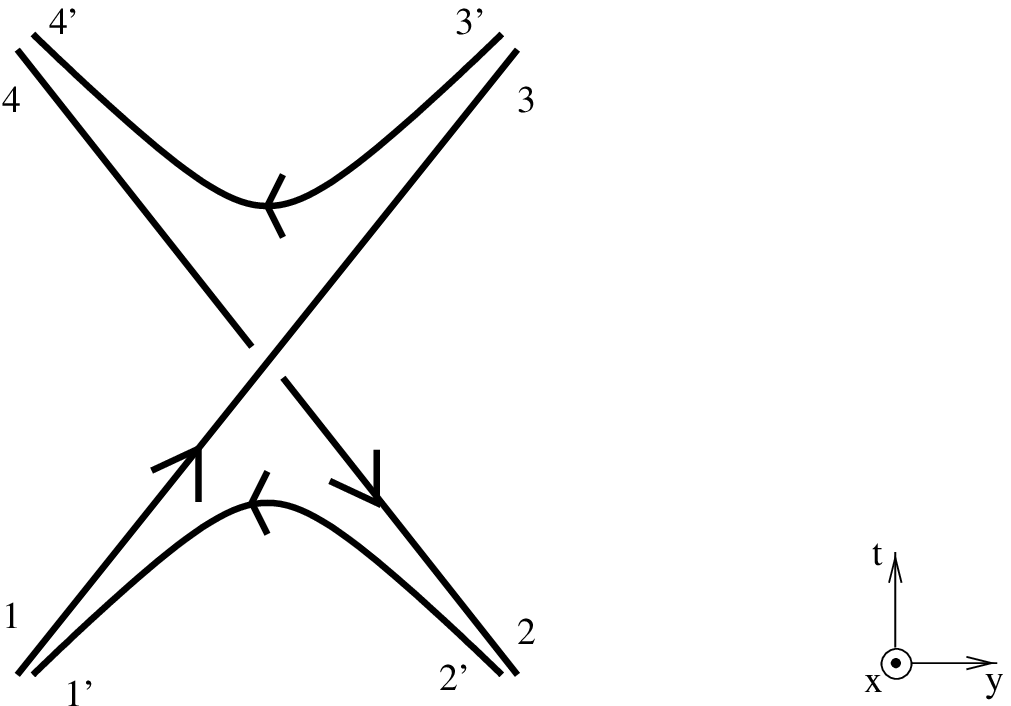} \hspace {2cm}
\includegraphics[width=3cm]{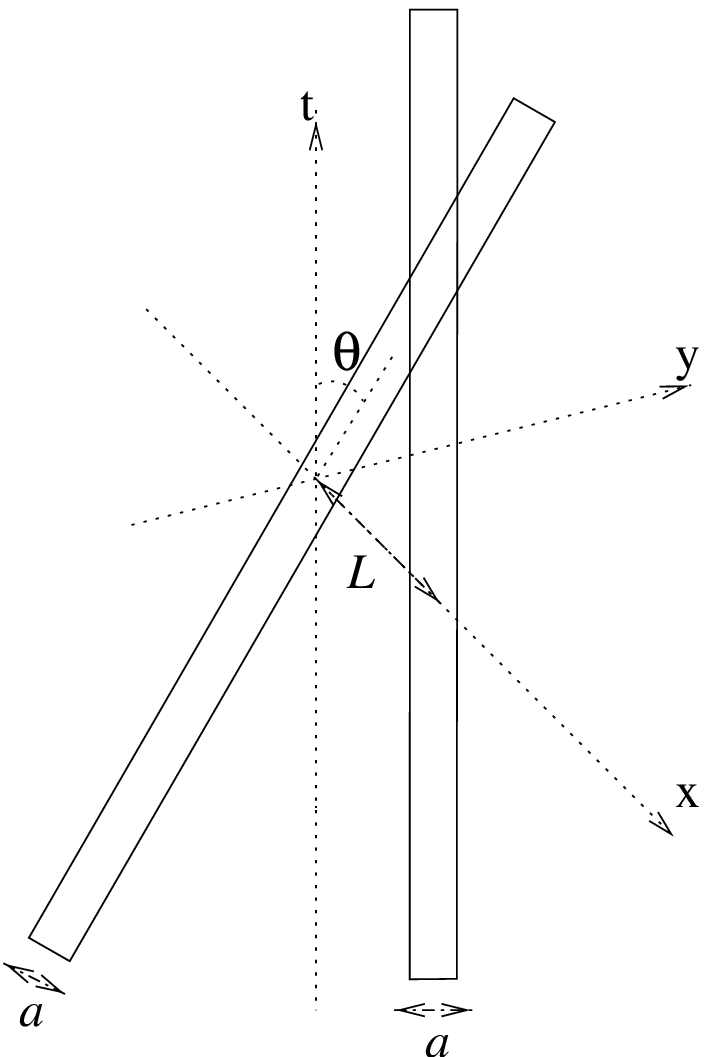}
\caption[]{\it Boundaries of Minimal surfaces for Reggeon and Pomeron 
amplitudes (from \cite{janik}).}
\label{reggeon}
\end{center}
\end{figure}

\subsubsection{The origin of ``Regge behavior'' in  non-pQCD}

 Even  a more formidable challenge is the origin of the Regge behavior in reactions 
where it has been displayed by data, \ie soft reactions corresponding to the 
unknown strong coupling regime of QCD. While phenomenological work may still be 
needed~\cite{vin}, new tools of handling non perturbative calculations for 
strongly coupled gauge theories have been investigated. Using the new duality 
properties relating strongly coupled gauge theories to weakly coupled string 
theories in a non-flat background space called $AdS/CFT$ duality, the Pomeron 
and Reggeon amplitudes have been related \cite{janik} to a particular 
geometrical problem: finding minimal surfaces corresponding to specific 
boundary  conditions, see Fig.~\ref{reggeon}. 
In this framework, ``Reggeization'', \ie linearly rising 
Regge trajectories, is related to the string background metrics in much the 
same way as confinement is related to the  Wilson loop area law. 

The obtained 
values of Pomeron and Reggeon intercepts and slopes  are in the correct range, 
but some  approximations in the calculation have to be improved to get more 
precise results.
The connection of $AdS/CFT$ duality with quantum field theory features (such as 
instantons, solitons) is highly desirable, the most difficult problem being the 
identification of the supergravity dual of QCD in a generalized AdS/CFT 
correspondence. Despite these difficulties, this new 
approach appears quite promising.

\section*{Acknowledgments}
This paper is dedicated to the memory of Bo Andersson, who died 
unexpectedly from a heart attack on March 4th, 2002. 
We all have learned so much from him. 
\par
We want to thank the organisers for this
nice and well prepared workshop and also all the participants of our working
group sessions for their contributions and the lively discussions.
\raggedright

\end{document}